\documentclass[aps,reprint,superscriptaddress,longbibliography]{revtex4-1}
\usepackage[colorlinks,bookmarks=false,citecolor=red,linkcolor=blue,urlcolor=blue]{hyperref}
\usepackage{amsmath,amssymb,graphicx,color}
\usepackage{booktabs}
\usepackage{bm} 
\usepackage{amssymb}
\usepackage{color}
\usepackage{multirow}
\usepackage{mathrsfs,bbm}
\usepackage{amsmath}
\usepackage{verbatim}
\usepackage{bbm}
\usepackage[version=3]{mhchem}
\usepackage{array}
\newcolumntype{P}[1]{>{\centering\arraybackslash}p{#1}}
\usepackage{epstopdf}
\usepackage{wrapfig}

\def\Hhat{\hat{H}}

\def\c{\hat{c}}
\def\a{\hat{a}}
\def\b{\hat{b}}

\def\Hhat{\hat{H}}

\def\veps{\varepsilon}

\def \r{{\bf r}}
\def \d{{\bm \delta}}

\def\up{\uparrow}
\def\down{\downarrow}

\definecolor{mycol}{rgb}{0,0,0}  
\begin{document}
\title{Multiterminal Josephson junctions with tunable topological properties}
\author{Panch Ram}
\email{panch.ram@uni-konstanz.de}
\affiliation{Fachbereich Physik, Universität Konstanz, D-78457 Konstanz, Germany}
\author{Detlef Beckmann}
\affiliation{Institute for Quantum Materials and Technologies, Karlsruhe Institute of Technology, Karlsruhe D-76021, Germany}
\author{Romain Danneau}
\affiliation{Institute for Quantum Materials and Technologies, Karlsruhe Institute of Technology, Karlsruhe D-76021, Germany}
\author{Wolfgang Belzig}
\email{wolfgang.belzig@uni-konstanz.de}
\affiliation{Fachbereich Physik, Universität Konstanz, D-78457 Konstanz, Germany}

\date{\today}	
\begin{abstract}
Since the discovery of the Andreev reflection process at normal-metal/superconductor junctions and the corresponding Andreev bound states in superconductor/normal-metal/superconductor junctions, various multiterminal Josephson junctions have been studied to explore many exotic phases of quantum matter, where the formation of Andreev bound states in the normal region account for dissipationless supercurrent and play a central role in determining exotic properties. Recently, an intriguing aspect of the multiterminal Josephson junctions has been proposed to study the topological properties, wherein the Andreev bound states acquire topological characteristics upon tuning the phase differences of superconducting terminals. In this work, we investigate topologically non-trivial phases in four-terminal Josephson junctions based on square and graphene lattices. Additionally, we apply a gating potential that smoothly drives the Andreev bound states from a topologically non-trivial state to a trivial state. Furthermore, we observe that the gating potential in our setup produces the similar physics of the topological Andreev bound states of the double (single) quantum-dot multiterminal Josephson junctions when the gating potential is small (large) compared to the superconducting gap.
\end{abstract}
\maketitle
\section{Introduction}
Although the underlying concepts of topology in mathematics have been known since the seventeenth century, the successful integration of this idea into condensed matter physics is relatively recent, emerging only about two decades ago with the discovery of novel topological quantum materials, such as topological insulators~\cite{Klitzing1980,Hatsugai1993,Kane2005,Hasan2010}, topological semimetals~\cite{Armitage2018}, and topological superconductors~\cite{Sato2017}. The topological insulators exhibit insulating bulk states while supporting conducting surface states at the edges, whereas the electronic states of the topological semimetals demonstrate novel responses to externally applied electric and magnetic fields. In the case of the topological superconductors, the time-reversal symmetry is broken in two-dimensional planar systems, and they differ fundamentally from the Bose-Einstein condensate of Cooper pairs. Moreover, the study of topological superconductors has advanced the development of Majorana fermions, which are their own antiparticles and obey non-Abelian braiding statistics~\cite{Alicea2012,Beenakker2013,Sato2016}. These Majorana zero-energy modes are particularly significant for their potential use as qubits in topological quantum computation~\cite{Sarma2015,Deng2016,Aasen2016,Karzig2017}.

The non-trivial topology in such real-world materials is often hard-coded, requiring some specific interactions such as strong spin-orbit coupling. However, a branch of topological investigations has recently gained attention in the condensed matter community--synthetic quantum matters--where internal degrees of freedom can be easily tuned, offering greater control over topological properties.  Examples of such synthetic quantum systems include topological photonics~\cite{Lu2014, Ozawa2019}, topological driven Floquet systems~\cite{Rudner2020}, topological electrical circuits~\cite{Imhof2018}, and multiterminal Josephson junctions (MJJs)~\cite{Riwar2016,Erik2017,Xie2017,Xie2018,Deb2018,Xie2019,Peralta2019,Houzet2019,Klees2020,Klees2021,Sept2023,Peralta2023,Riwar2019,Javed2023}. The MJJs, in particular, comprise multiple BCS-type superconducting leads and can, in principle, host topologically non-trivial Andreev bound states (ABSs) in synthetic space of the superconducting phase differences. These topological ABSs result in an integer-valued first Chern number which is related to a quantized transconductance between two superconducting terminals~\cite{Riwar2016}. Moreover, in the MJJs, a non-trivial higher-dimensional topology can be easily engineered by simply increasing the number of superconducting leads~\cite{Riwar2016,Weisbrich2021}.

Recently, various types of multiterminal Josephson junctions have been synthesized and subjected to experimental measurements; however, a clear signature of the non-trivial topology in such systems remains elusive, due to experimental challenges in achieving the specific conditions required by the scattering region placed in between the superconducting terminals~\cite{Draelos2019,Pankratova2020,Arnault2021,Chandrasekhar2022,Coraiola2023,Prosko2024,Wisne2024}. Alternatively, several proposals involving quantum dots coupled with superconducting terminals have been suggested, but these also require specialized coupling between the superconducting terminals, which presents further feasibility limitations~\cite{Klees2020,Klees2021,Sept2023}. More recently, however, a system consisting of a double quantum-dot coupled with four superconducting terminals has been proposed, which may be experimentally feasible~\cite{Lev2023}. In this work, we investigate the topological properties of Andreev bound states in four-terminal Josephson junctions based on square and graphene lattices. Graphene is ideally suited for experiments on multiterminal Josephson junctions due to the high quality of graphene encapsulated in boron nitride, the gate tunability, and the possibility of forming high-transparency contacts with superconductors~\cite{Calado2015,Benshalom2016,Borzenets2016,Nanda2017,Kraft2018,Zhu2018,ParkPRL2018,Schmidt2018,Pandey2019,Pandey2021,Pandey2022,Schmidt2023,Messelot2024}. Ballistic transport is routinely observed, and the limit of short Josephson junctions can be easily achieved in hybrid structures defined by e-beam lithography. Nonetheless, real-world structures will be much larger than a single quantum dot and incorporate inhomogeneous gate potentials. We therefore focus here on finite-size systems with gate potential included. In the short-junction scattering limit, we demonstrate that the lowest ABSs undergo a topological phase-transition upon tuning the superconducting phase differences between the terminals. In addition, we examine the robustness of these topological ABSs in the presence of an externally applied gate potential in the scattering region. We find that the topological nature of these ABSs remains intact for a large gate potential relative to the superconducting energy gap, before eventually transitioning into a topologically trivial phase.

The rest of the paper is ogranized as follows. In Sec.~\ref{sec:model-method}, we present the tight-binding model Hamiltonian of MJJs for both square and graphene lattices, the scattering matrix theory formalism to obtain the ABS energies and the corresponding states, and a numerical method~\cite{Fukui2005} to calculate the Chern number. Sec.~\ref{sec:result} provides results for the topological phase diagrams, obtained by using the Chern number and minima of the lowest positive ABS energy, in the space of model parameters. Finally, we conclude the paper in Sec.~\ref{sec:conclusion}.
\section{Model and Method}\label{sec:model-method}
\subsection{Hamiltonian}
\begin{figure}[htb]
        \includegraphics[width=0.48\textwidth]{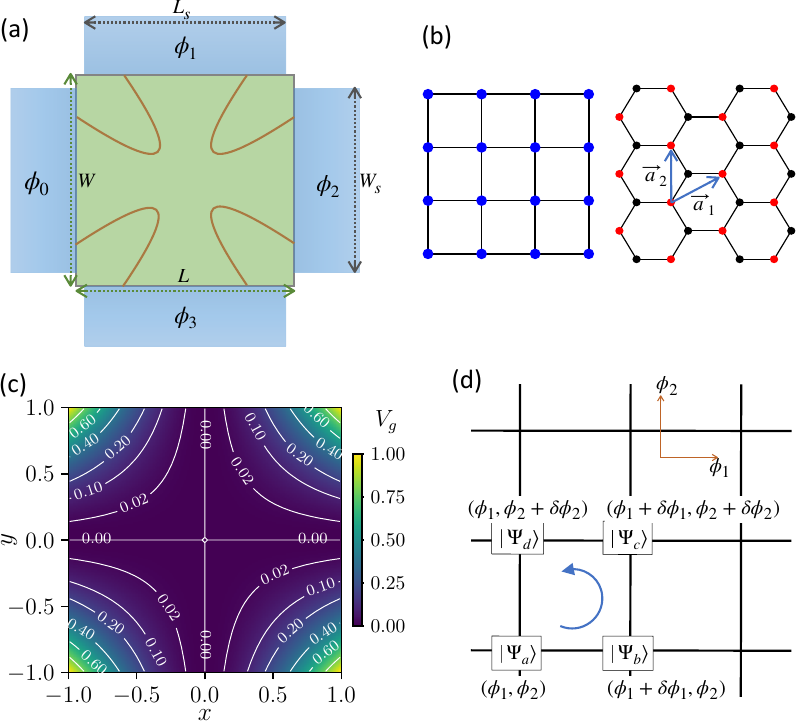}
	\caption{(a) Schematic illustration of the multiterminal Josephson junctions setup. We consider a system with four superconducting terminals attached with a scattering region, forming Josephson junctions. The scattering region is indicated in light-green with length $L$ and width $W$ whereas the four superconducting (SC) leads are depicted in the light-blue with phases $\phi_\alpha$ (for $\alpha=0$ to $3$). These semi-infinite horizontal and vertical leads have the corresponding width $W_s$ and length $L_s$ respectively. Additionally, a schematic of externally applied gating potential, acting only in the scattering region, is shown in orange color, increasing in strength along the diagonals. (b) We setup the system based on square and graphene lattices and employ the KWANT~\cite{Groth2014} to numerically simulate the tight-binding model Hamiltonians in Eqs.~\eqref{eq:hamilt-sq} and~\eqref{eq:hamilt-gr}. (c) A typical color plot and contour plot for the applied gating potential of the form $V(x,y) = V_g x^2 y^2$.
    (d) A discretized mesh-grid in ($\phi_1, \phi_2$) space is used to numerically calculate the Chern number, in Eq.~\eqref{eq:chern-discrete}, by using the Fukui, Hatsugai, and Suzuki method~\cite{Fukui2005}.}
	\label{fig:system}
\end{figure}

We setup a tight-binding model Hamiltonian for a system of four-terminal Josephson junctions, as shown in Fig.~\ref{fig:system}, based on square and graphene lattices to study the topologically non-trivial Andreev bound states, $\Hhat = \Hhat_0 + \Hhat_\Delta$. For the square lattice:
\begin{subequations}
    \begin{align}
    \Hhat_0 = &  - t\sum_{\r, \d} \sum_{\sigma=\up,\down} ( \c_{\r, \sigma}^\dag \c_{\r+\d, \sigma}  + \mathrm{h.c.} ) \nonumber\\
    &+  \sum_\r \sum_{\sigma=\up,\down}   \left( 4t -\mu + V_\r \right)  \c_{\r, \sigma}^\dag  \c_{\r, \sigma} \\
    \Hhat_\Delta = & \sum_\r ( \Delta_\r \c_{\r, \up}^\dag \c_{\r, \down}^\dag + \Delta_\r^{*} \c_{\r, \down} \c_{\r, \up} )
    \end{align}
    \label{eq:hamilt-sq}
\end{subequations}
where the operator $\c_{\r,\sigma}^\dag (\c_{\r,\sigma})$ creates (annihilates) an electron with spin $\sigma\in \{\up,\down\}$ at lattice site $\r\equiv n_x \hat{x} + n_y \hat{y}$ with $n_x, n_y\in \mathbb{Z}$ and $\d\equiv \{\pm \hat{x}, \pm \hat{y}$\} is the nearest-neighbor vector, $t$ and $\mu$ are respectively the hopping energy amplitude and chemical potential, while $V_\r\equiv V(x,y)= V_g x^2y^2$ is added to account for the gating potential which is only applied in the scattering region. The superconducting pairing $\Delta_\r=\Delta e^{i\phi_\alpha}$ is non-zero only for the lattice sites that lie in the superconducting leads, which are labeled for the index $\alpha\in \{0,1,2,3\}$ with the corresponding phases $\{\phi_\alpha\}$, and accounts for the $s$-wave singlet superconducting pairing. The constant $4t$ onsite energy term in $\Hhat_0$ is added to shift the overall energy spectrum.

Similarly, for the graphene lattice:
\begin{subequations}
\begin{align}
	\hat H_0 =& -t \sum_{\r,\d}\sum_{\sigma=\up,\down}  ( \a_{\r+\d,\sigma}^\dag \b_{\r,\sigma} + \text{h.c.} ) \nonumber \\ 
	&+ \sum_\r \sum_{\sigma=\up,\down}  (V_\r - \mu)  ( \a_{\r,\sigma}^\dag \a_{\r,\sigma} + \b_{\r,\sigma}^\dag \b_{\r,\sigma} ) \\
	\hat H_\Delta =& \sum_\r \left[ (\Delta_\r \a_{\r,\up}^\dag \a_{\r,\down}^\dag + \Delta_{\r+\bm \tau} \b_{\r,\up}^\dag \b_{\r,\down}^\dag) + \text{h.c.} \right]
\end{align}
	\label{eq:hamilt-gr}
\end{subequations}
where $\r\equiv n_1 \vec{a}_1 + n_2\vec{a}_2$ is lattice vector with integers $n_1, n_2\in \mathbb{Z}$ and $\vec{a}_1=a(\sqrt{3}/2, 1/2)$ and $\vec{a}_2=a(0, 1)$ are the primitive vectors for graphene with lattice constant $a$. Here, we separately define the creation (annihilation) operators $\a_{\r,\sigma}^\dag$ and $\b_{\r,\sigma}^\dag$ ($\a_{\r,\sigma}$ and $\b_{\r,\sigma}$) with spin $\sigma\in \{\up,\down\}$ on two sublattices A and B, respectively, at the lattice site positions $\r$ and $\r +{\bm\tau}$ with ${\bm\tau} = a\hat{x}/\sqrt{3}$, see in Fig.~\ref{fig:system}(b). However, we consider the same $s$-wave superconducting pairing $\Delta_\r=\Delta_{\r+\bm \tau} \equiv \Delta e^{i\phi_\alpha}$ on both sublattices.
\subsection{Scattering matrix theory approach}
To calculate the topological Andreev bound states, we use KWANT software~\cite{Groth2014} to set up the tight-binding model Hamiltonians, given in Eqs.~\eqref{eq:hamilt-sq} and~\eqref{eq:hamilt-gr} for square and graphene lattices, and employ the scattering matrix theory approach, which is described thoroughly in Ref.~\cite{Heck1024} for the use of multiterminal Josephson junctions. This approach is well suited for a short junction limit, i.e., when the length ($L$) of the scattering region is much smaller than the superconducting coherence length ($\xi$). In this limit, the corresponding ABS energies, $|\veps|<\Delta$, formed due to electron-hole conversion process at the all interfaces, are determined by the condition,
\begin{align}
	S_A(\veps)S_N(\veps)|\Psi\rangle = |\Psi\rangle
\end{align}
which involves the scattering matrices at the interfaces for Andreev reflection process ($S_A$) as well as normal refection process ($S_N$). Within this approach, their forms can be written as:
\begin{align}
	S_A(\veps) = \beta(\veps)
	\begin{pmatrix}
		0 & r_A^* \\[0.3em]
		r_A& 0
	\end{pmatrix},~
	S_N(\veps) =
	\begin{pmatrix}
		S(\veps)& 0 \\[0.3em]
		0& S^*(-\veps)
	\end{pmatrix}
\end{align}
with $\beta(\veps)=\sqrt{1-(\veps/\Delta)^2} + i(\veps/\Delta)$ and the Andreev reflection matrix $r_A$ is in diagonal form,
\begin{align}
	r_A = \begin{pmatrix}
		ie^{i\phi_0}\mathbf{1}_{n_0}\\[0.3em]
		~& ie^{i\phi_1}\mathbf{1}_{n_1} \\[0.3em]
		~& ~& ie^{i\phi_2}\mathbf{1}_{n_2} \\[0.3em]
		~& ~& ~& ie^{i\phi_3}\mathbf{1}_{n_3}
	\end{pmatrix}
\end{align}
where $\{\mathbf{1}_{n_\alpha}\}$ are the identity matrices for the incoming scattering modes $\{n_\alpha\}$ in the leads $\alpha=0$ to $3$, whereas $S(\veps) (S^*(-\veps))$ corresponds to the electron (hole) scattering matrix-block. Utilizing the short-junction limit approximation, i.e., $S(\veps)\approx S(\veps=0)\equiv s$, we arrive at the following eigenvalue-equation for the ABS states,
\begin{align}
		\begin{pmatrix}
		s^\dag & 0 \\[0.3em]
		0& s^T
	\end{pmatrix}
	\begin{pmatrix}
	0 & r_A^* \\[0.3em]
	r_A& 0
	\end{pmatrix} |\Psi\rangle = \beta(\veps)|\Psi\rangle.
\label{eq:scatt-ABS}
\end{align}
Now, solving the above Eq.~\eqref{eq:scatt-ABS} yields the ABS energies and the corresponding eigenstates.
\subsection{Chern number}
In our four-terminal Josephson junctions setup, the ABS energies and eigenstates ($\veps_n, |n\rangle$ for the index $n=\pm1,\pm2,\cdots$) depend on the four superconducting phases. Since out of the four phases only three are independent, we use a gauge variance to fix one phase $\phi_0=0$ and determine the ABS energies $|\veps_n|<\Delta$ and wavefunctions with respect to the remaining phases $\phi_\alpha$ (for $\alpha=1,2,3$)~\cite{Klees2020,Klees2021}. These phases $(\phi_1, \phi_2, \phi_3)\in [0, 2\pi)^3$ form a three-dimensional periodic compact space analogous to the quasi-momenta in periodic crystals. In our setup, the lowest positive ABS energies, $\veps_{n}$ for $n=\pm 1$, close and open the gap at zero energy upon tuning the phase differences. These phases act as the quasi-momenta in this synthetic dimensions for the topological Andreev bound states. Therefore, for a given phase difference, such as $\phi_3$, we define the $n$th state Chern number in the space of other two phase differences ($\phi_1, \phi_2$) as,
\begin{align}
C_{12}^{(n)} = \frac{1}{2\pi} \int_{0}^{2\pi}\int_{0}^{2\pi} B_{12}^{(n)} d\phi_1 d\phi_2
\label{eq:chern}
\end{align}
where $B_{12}^{(n)}=\partial_1 A_2^{(n)}-\partial_2 A_1^{(n)}$ is the Berry curvature for the $n$th state, while $A_{1/2}^{(n)}=i\langle n|\partial_{1/2}|n\rangle$ is the corresponding Berry connection. The $C_{12}^{(n)}$ is quantized to integer values, which leads to a quantized transconductance $G_{12} = (-4e^2/h) C_{12}^{\mathrm{GS}}$ between the terminals $1$ and $2$. Here $h$ is the Planck constant, $e$ is the elementary charge, and $C_{12}^{\mathrm{GS}}$ is the ground-state Chern number at zero temperature, obtained by summing over all negative-energy eigenstates, i.e., $C_{12}^{\mathrm{GS}} = \sum_n C_{12}^{(n)}$ such that $\veps_n<0$.
\begin{figure*}[htb]
	\includegraphics[width=0.975\textwidth]{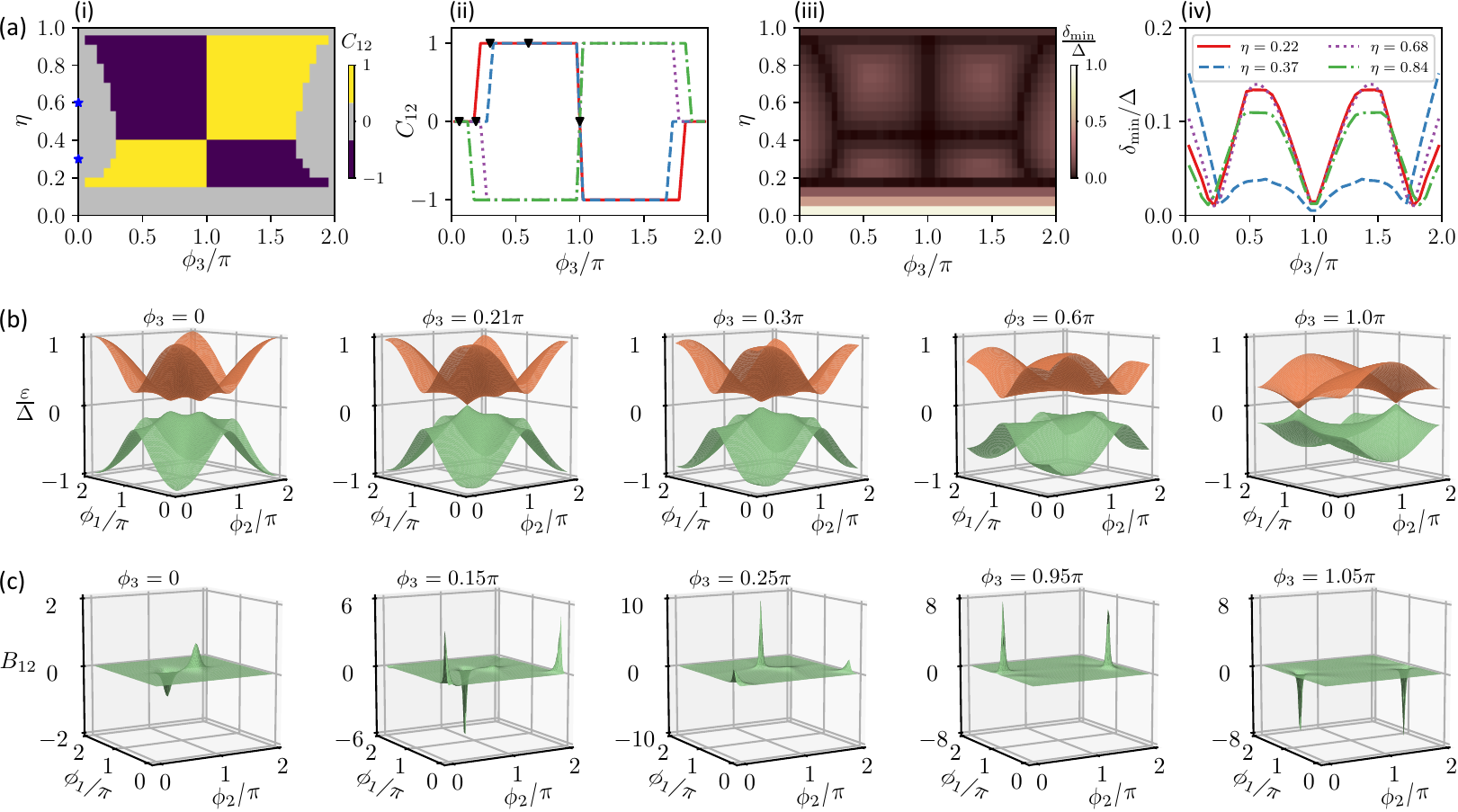}
	\caption{(a) Topological phase diagram for the Chern number $C_{12}^{\mathrm{GS}}$ in (i) and the corresponding phase boundaries from minimum gap $\delta_{\mathrm{min}}$ (of the lowest ABS energy) in (iii) are plotted in the parameters space $(\phi_3, \eta)$ when the gating potential is set to zero, i.e., $V_g=0$; whereas, they are also shown for a few selected values of $\eta$ in (ii) and (iv). (b)-(c) At a fixed $\eta=0.22$, the Andreev-energy dispersions $\veps_{\pm1}$ (in unit of $\Delta$) and the Berry curvature $B_{12}$ (for the lower band) with respect to the phase differences $\phi_1$ and $\phi_2$ for a set of different $\phi_3$ values. The chosen $\phi_3$ values are shown in the corresponding plots of (b) and (c); while particularly for the $\veps_{\pm1}$, they are also shown in (a)-(ii) as inverted black triangles.}
	\label{fig:chern-zerogate}
\end{figure*}

As we only obtain numerical data for the eigenspectrum of Eq.~\eqref{eq:scatt-ABS}, within the KWANT simulation, we use the numerical method developed by Fukui, Hatsugai, and Suzuki in Ref.~\cite{Fukui2005} to calculate the Chern number. Specifically, we discretize the phases $0<\phi_{1/2} < 2\pi$ into $M\times M$ grid-points, with finite difference $\delta\phi_{1/2} = 2\pi/M$, and define the link quantity $U_{\delta\vec{\phi}}^{(n)}(\vec{\phi}) = \langle n(\vec{\phi})|n(\vec{\phi} + \delta\vec{\phi})\rangle$. Here, $\vec{\phi}$ is a grid point vector in the ($\phi_1, \phi_2$) space, and $\delta\vec{\phi}$ represents the step to the next grid point. Then, analogous Berry curvature is evaluated in terms of link quantities as,
 \begin{align}
 	\tilde{B}_{12}^{(n)}(\vec{\phi})= \log \big[&U_{\delta\phi_1}^{(n)}(\vec{\phi}) U_{\delta\phi_2}^{(n)}(\vec{\phi} + \delta\phi_1) \nonumber\\ 
 	&\times U_{\delta\phi_1}^{(n)^{-1}}(\vec{\phi} + \delta\phi_2)  U_{\delta\phi_2}^{(n)^{-1}}(\vec{\phi}) \big].
 	\label{eq:analogous-B}
 \end{align}
To better understand these link quantities, we consider the $n$th eigenstates $|\Psi_a\rangle$, $|\Psi_b\rangle$, $|\Psi_c\rangle$, and $|\Psi_d\rangle$ for a plaquette [see in Fig.~\ref{fig:system}(d)] at the grid points ($\phi_1, \phi_2$), ($\phi_1+\delta\phi_1, \phi_2$), ($\phi_1+\delta\phi_1, \phi_2+\delta\phi_2$), and ($\phi_1, \phi_2+\delta\phi_2$), respectively. Then the analogous Berry curvature in Eq.~\eqref{eq:analogous-B} is expressed as,
  \begin{align}
 	\tilde{B}_{12}^{(n)}(\vec{\phi})= \log\big[ \langle \Psi_a|\Psi_b\rangle  \langle \Psi_b|\Psi_c\rangle  \langle \Psi_c|\Psi_d\rangle  \langle \Psi_d|\Psi_a\rangle  \big].
 \end{align}
Finally, the Chern number is obtained by summing over all $M\times M$ grid points as: 
 \begin{align}
C_{12}^{(n)} = \frac{1}{2\pi} \sum_{\{\vec{\phi}\}}\Im[\tilde{B}_{12}^{(n)}(\vec{\phi})]
\label{eq:chern-discrete}
\end{align}
where the symbol $\Im$ denotes the imaginary part. Note that $\tilde{B}_{12}^{(n)}$ and $B_{12}^{(n)}$ in Eqs.~\eqref{eq:analogous-B} and~\eqref{eq:chern} are closely related, with $\tilde{B}_{12}^{(n)}(\vec{\phi}) \simeq B_{12}^{(n)}(\vec{\phi})\delta\phi_1 \delta\phi_2$.

In the next section, we present the results for the topological phase diagrams of the Chern number $C_{12}^{\mathrm{GS}}$ and the $\delta_{\mathrm{min}}\equiv \mathrm{min}_{(\phi_1, \phi_2)}\veps_1$ of lowest Andreev band for both square and graphene lattices. Additionally, we show the evolution of ABS energies $\veps_n$ and Berry curvature $B_{12}^{(n)}$ with respect to $(\phi_1, \phi_2)$ for different values of $\phi_3$. These quantities are calculated using Eqs.~\eqref{eq:scatt-ABS},~\eqref{eq:analogous-B}, and~\eqref{eq:chern-discrete}. For the calculation, we set $t=1$, with all energy parameters expressed in units of $t$, and fix the superconducting pairing value to $\Delta=0.005t$.
\section{Results}\label{sec:result}
\subsection{Square}\label{subsec:square}
For the square lattice, we consider the system size of our four-terminal junctions shown in Fig.~\ref{fig:system}(a)-(b), with the scattering region $L=W=24a$, whereas the widths of semi-infinite leads $0$ and $2$ have $W_s=L-2$, and the lengths of leads $1$ and $3$ have $L_s=W_s$. We parameterize the global chemical potential as $\mu=(1-\eta)E_1 + \eta E_2$, where $\eta \in (0,1)$ is an independent parameter. The values of $E_1$ and $E_2$ are fixed using the dispersion relation formula for a square lattice, $E_{n}(k) = 4t - 2t\cos(ka) - 2t\cos(\frac{\pi a}{W_s}n)$, of an ideal semi-infinite sheet of width $W_s$. We choose the two lowest mode energies $E_1 \equiv E_{n=1}(k=0) \approx 4.07\Delta$ and $E_2 \equiv E_{n=2}(k=0)\approx 16.2\Delta$ and set $E_1<\mu<E_2$ for $\eta \in (0,1)$. This condition ensures that only one conducting channel is available in each lead for the Andreev process of electron-hole conversion.

We first set the gating potential to zero (i.e., $V_g=0$) and present results for the non-trivial topological phases in terms of the ground state Chern number, $C_{12}^{\mathrm{GS}}$, calculated from Eq.~\eqref{eq:chern-discrete}, and also the corresponding phase boundaries are obtained by evaluating the minimum of the lowest ABS energy $\delta_{\mathrm{min}}\equiv \mathrm{min}_{(\phi_1, \phi_2)}\veps_1$, from Eq.~\eqref{eq:scatt-ABS}, in the parameter space of $\phi_3$ and $\eta$. Figure~\ref{fig:chern-zerogate}(a)(i) shows the phase diagram for the Chern number, highlighting large stable regions (yellow and dark purple) where $C_{12}^{\mathrm{GS}}\ne 0$, indicating topologically non-trivial phases, while the gray region, where $C_{12}^{\mathrm{GS}}=0$, corresponds to a topologically trivial phase. We observe a phase transition from the topologically trivial phase to the non-trivial phase along both the $\phi_3$ and $\eta$ directions. We obtain results similar to the double quantum-dot system attached to four superconducting leads when varing $\phi_3$~\cite{Lev2023}, showing a non-trivial topological phase transition from $C_{12}^{\mathrm{GS}}=1(-1)$ to $C_{12}^{\mathrm{GS}}=-1(1)$, depending on $\eta$, at $\phi_3=\pi$. Interestingly, by varying $\eta$, i.e., the global chemical potential $\mu$, the non-trivial Chern number also changes around $\eta\approx 0.4$ even though the conducting channel for the Andreev process remains fixed. Additionally, we determine the phase boundaries by simply tracking the minimum value of the lowest positive ABS energy, see in Fig.~\ref{fig:chern-zerogate}(a)(iii). It clearly marks the boundaries where the gap becomes zero, i.e., $\delta_{\mathrm{min}}=0$. Several plots of $C_{12}^{\mathrm{GS}}$ and $\delta_{\mathrm{min}}$ are explicitly shown in Figs.~\ref{fig:chern-zerogate}(a)(ii) and~\ref{fig:chern-zerogate}(a)(iv) as a function of $\phi_3$ for selected values of $\eta= 0.22, 0.37, 0.68$, and $0.84$. Clearly, as $\eta$ increases, the topologically non-trivial phases for $C_{12}^{\mathrm{GS}}$ initially shrink, but after $\eta\approx 0.4$, these regions expand again. However, with further increase in $\eta$, the system eventually enters the topologically trivial-phase, see also Fig.~\ref{fig:chern-zerogate}(a)(i).

To better understand these phases, we plot the subgap Andreev energy bands ($\veps_{\pm 1}$) as functions of $\phi_1$ and $\phi_2$, in Fig.~\ref{fig:chern-zerogate}(b), at $\eta=0.22$ for various values of $\phi_3$ [shown also in Fig.~\ref{fig:chern-zerogate}(a)(ii) as inverted-triangle]. We observe that the overall Andreev spectrum remains gapped at $\phi_3=0$. However, as $\phi_3$ increases, both bands touch at zero energy and become gapless at a certain critical value of $\phi_3\approx 0.21\pi$ at a single point (Weyl point). Further tuning of $\phi_3$ leads to reopening of the gap and the system goes into a topologically non-trivial regime with $C_{12}^{\mathrm{GS}}=1$, see the subplots for $\phi_3=0.3\pi$ and $0.6\pi$. The gap closes again at $\phi_3=\pi$, now at two Weyl points, and opens with increasing $\phi_3>\pi$, exhibiting another topological phase transition with $C_{12}^{\mathrm{GS}}=-1$ (the Andreev bands for $\pi<\phi_3\leq 2\pi$ are not plotted for brevity, as they look similar to the plots for $0\leq\phi_3\leq\pi$). In Fig.~\ref{fig:chern-zerogate}(c), we present the Berry curvature $B_{12}$ for the lower Andreev band $\veps_{n=-1}$ at $\eta=0.22$ for a set of selected values of $\phi_3= 0, 0.15\pi, 0.25\pi, 0.95\pi, 1.05\pi$. It is evident that $B_{12}$ changes sign at the Weyl point(s) near the critical values of $\phi_3$ where the phase transition occurs, for instance, see the second and third subplots of $B_{12}$ near the $\phi_3=0.21\pi$ and the last two subplots near $\phi_3=\pi$ in Fig.~\ref{fig:chern-zerogate}(c). This demonstrates that the Andreev states exhibit a topologically nontrivial nature in the parameter space of the superconducting phase differences and chemical potential.
\begin{figure}[htb]
	\includegraphics[width=0.48\textwidth]{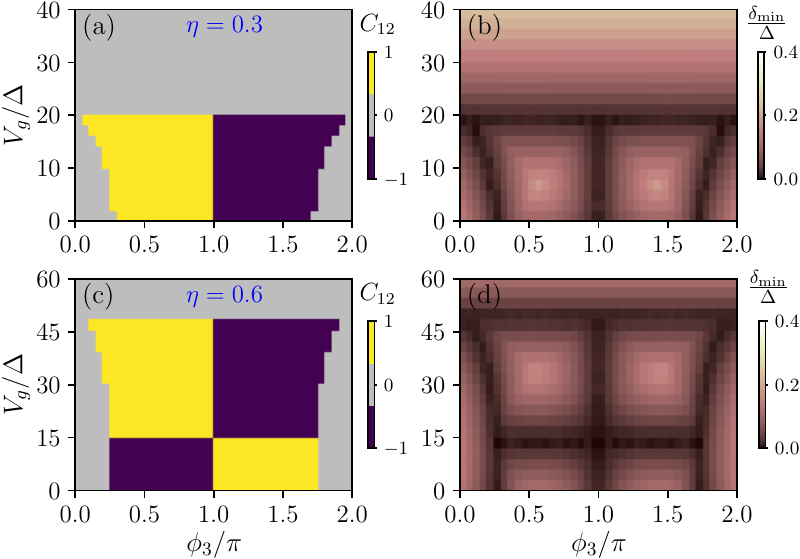}
	\caption{Topological phases for Chern number $C_{12}^{\mathrm{GS}}$ and the corresponding phase boundaries from $\delta_{\mathrm{min}}$ with respect to $\phi_3$ and $V_g$ for fixed values of $\eta=0.3$ and $0.6$.}
	\label{fig:chern-sq-vg}
\end{figure}

Next, we introduce the gating potential (i.e., $V_g\ne0$) and investigate its effect on the topologically non-trivial phases while keeping the global chemical potential fixed at $\eta=0.3$ and $0.6$, respectively. Both points (blue star markers) are indicated in Fig.~\ref{fig:chern-zerogate}(a)(i). In Fig.~\ref{fig:chern-sq-vg}, we show the results for $C_{12}^{\mathrm{GS}}$ and $\delta_{\mathrm{min}}$ in the parameter space of $\phi_3$ and $V _g$. For $\eta=0.3$, the stable topologically non-trivial regions expand with increasing strength of $V_g$, and the Chern number becomes $C_{12}^{\mathrm{GS}}=\pm 1$ for almost the entire range of $\phi_3$, before eventually transitioning into the topologically trivial phase with $C_{12}^{\mathrm{GS}}=0$, see in Fig.~\ref{fig:chern-sq-vg}(a). These phase transition boundaries are again accurately captured by the condition $\delta_{\mathrm{min}}=0$, as shown in Fig.~\ref{fig:chern-sq-vg}(b). Interestingly, for $\eta=0.6$ in Fig.~\ref{fig:chern-sq-vg}(c), we observe that tuning $V_g$ the non-trivial phase transition changes from $C_{12}^{\mathrm{GS}}=-1(1)$ to $C_{12}^{\mathrm{GS}}=1(-1)$ well before the system transitions into the trivial phase. The minima of the Andreev lower band again show these phase boundaries, see in Fig.~\ref{fig:chern-sq-vg}(d). Overall, we find that the topologically non-trivial phases remain robust even when the gating potential is turned on. Remarkably, $V_g$ enhances the stable regions of topologically non-trivial phases for moderate to large values compared to the superconducting energy gap.
\subsection{Graphene}\label{subsec:graphene}
For the graphene lattice, we set up the four-terminal system as shown in Figs.~\ref{fig:system}(a) and~\ref{fig:system}(c). The system size for the scattering region is $L=20a$ and $W=12a$; however, $W_s=W$ for the leads $0$ and $2$, while $L_s=W_s$ for the leads $1$ and $3$, respectively. Similar to the square lattice, we parametrize the global chemical potential as $\mu=(1-\eta)E_1 + \eta E_2$, where $E_1$ and $E_2$ are fixed using the conduction band dispersion relation: $E_{n}(k) = t\sqrt{1+4\cos(\frac{ka}{2})\cos(\frac{\pi a}{W_s}n) + \cos^2(\frac{\pi a}{W_s}n) }$ for an ideal semi-infinite sheet of armchair graphene with width $W_s$. We fix $E_1<\mu<E_2$ for $\eta\in(0,1)$ by choosing the two lowest energy modes $E_1 \equiv E_{n=4W_s/3}(k=0)=0$ and $E_2 \equiv E_{n=4W_s/3+1}(k=0)\approx 21.7\Delta$, respectively.
\begin{figure}[htb]
	\includegraphics[width=0.48\textwidth]{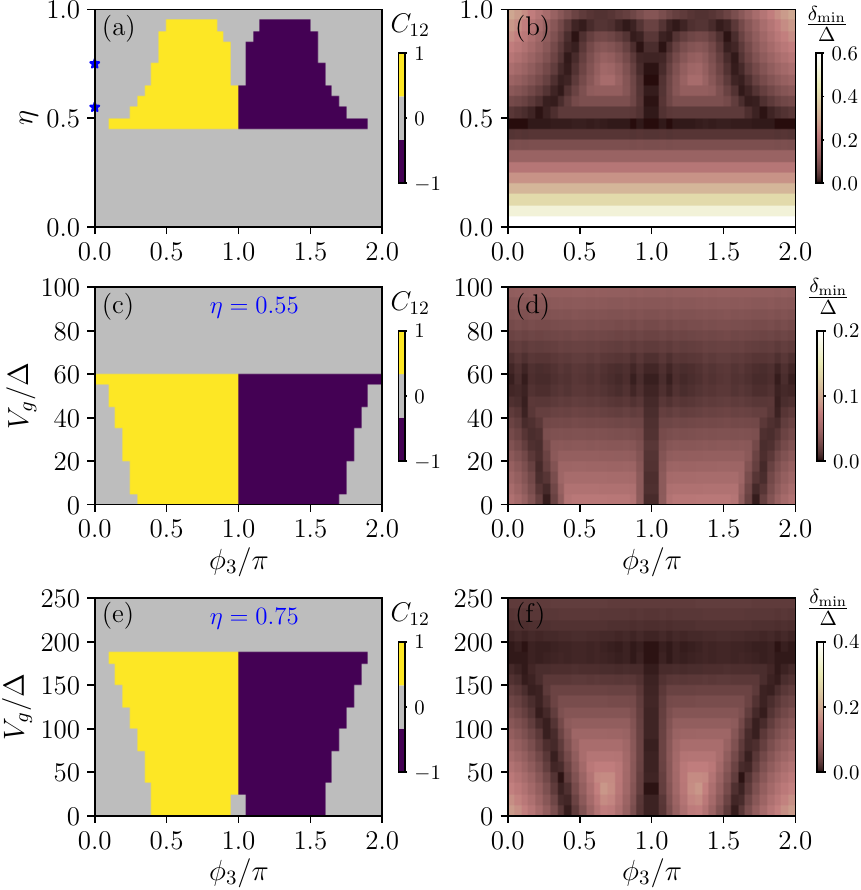}
	\caption{(a)-(b)Topological non-trivial phases for the graphene lattice, obtained from the Chern number $C_{12}^{\mathrm{GS}}$ and corresponding phase boundaries from $\delta_{\mathrm{min}}$, in parameters space of $\phi_3$ and $\eta$ when the gating potential is absent $V_g=0$. For fixed $\eta=0.55$ in (c)-(d) and $\eta=0.75$ in (e)-(f), the topologically non-trivial phases are robust and evolve upon tuning $V_g$.}
	\label{fig:chern-graphene}
\end{figure}

Now, we present the results for the topological phase diagram of the graphene based four-terminal system. In Figs.~\ref{fig:chern-graphene}(a) and~\ref{fig:chern-graphene}(b), we plot $C_{12}^{\mathrm{GS}}$ and $\delta_{\mathrm{min}}$, respectively, in the parameter space ($\phi_3, \eta$) for $V_g=0$. Notice that initially the system remains in the topologically trivial phase with $C_{12}^{\mathrm{GS}}=0$ as $\eta$ is tuned. However, it transitions into the topologically non-trivial phases with $C_{12}^{\mathrm{GS}}=\pm1$ upon further tuning. Eventually, it re-enters the trivial phase as $\eta$ approaches $1$. Additionally, unlike the square lattice, we observe a bifurcation around the phase transition point $\phi_3=\pi$ of the topologically non-trivial phases. These phase boundaries are also clearly observed in Fig.~\ref{fig:chern-graphene}(b) where $\delta_{\mathrm{min}}$ becomes zero. Next, we show the results in the presence of $V_g$ for fixed $\eta$ in the parameter space ($\phi_3, V_g$). For $\eta=0.55$ in Figs.~\ref{fig:chern-graphene}(c)-(d) we observe a phase diagram similar to that in Figs.~\ref{fig:chern-sq-vg}(a)-(b) for the square lattice. Furthermore, when we set $\eta=0.75$, the bifurcation around $\phi_3=\pi$ closes as $V_g$ increases. In Figs.~\ref{fig:chern-graphene}(e)-(f), we notably do not observe the phase transition between the topologically non-trivial phases upon tuning $V_g$, unlike the case in the square lattice in Figs.~\ref{fig:chern-sq-vg}(c)-(d). Nevertheless, it is present along $\phi_3$, which is the common feature of both lattices.
\subsection{Bogoliubov-de Gennes method}
In this subsection, we validate the results obtained from the scattering matrix theory approach in the previous two subsections using the Bogoliubov-de Gennes (BdG) method. We now consider a system with the superconducting leads of finite length attached to the scattering region, as shown in Fig.~\ref{fig:system}(a), and set up the Bogoliubov-de Gennes Hamiltonian,
\begin{align}
	H_{\mathrm{BdG}} = 
	\begin{pmatrix}
		H_0 & \Delta_\r  \\[0.3em]
		\Delta_\r^* & -H_0
	\end{pmatrix}
	\label{eq:DbG}
\end{align}
for both the lattices in the KWANT~\cite{Groth2014}. The Hamiltonain $H_{\mathrm{BdG}}$ is written in the particle-hole basis using the model Hamiltonians given in Eqs.~\eqref{eq:hamilt-sq} and~\eqref{eq:hamilt-gr} for square and graphene lattices. We numerically compute the full energy spectrum by diagonalizing $H_{\mathrm{BdG}}$ Hamiltonian matrix in Eq.~\eqref{eq:DbG} and sort the Andreev subgap energy states with $|\veps|<\Delta$ to calculate the topological Chern number. The length of the SC leads is set to $L_{SC} \simeq 12L$ to ensure the short junction limit for $\Delta=0.005t$. We use a standard sparse matrix technique to numerically diagonalize the large $H_{\mathrm{BdG}}$ matrix. The system size and model parameters are considered the same as those used in subsections~\ref{subsec:square} and~\ref{subsec:graphene} for the scattering matrix (SM) method for square and graphene lattices.
\begin{figure}[htb]
	\includegraphics[width=0.48\textwidth]{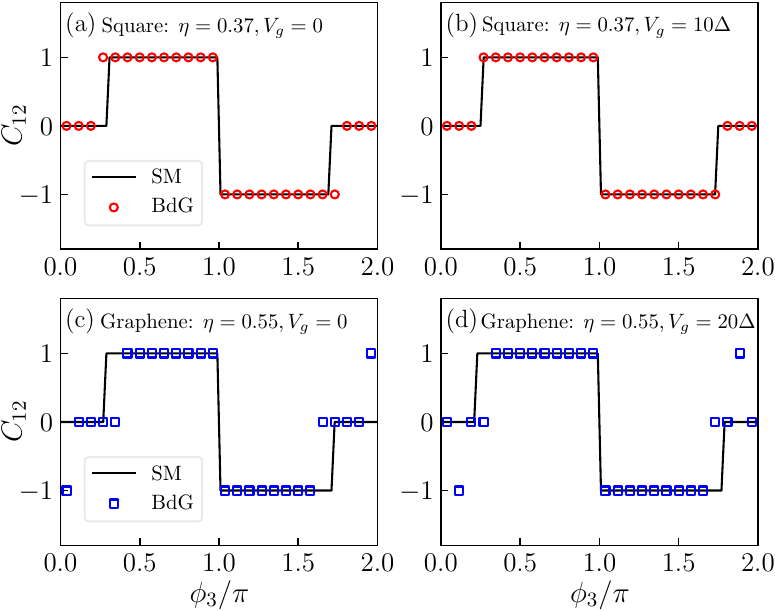}
	\caption{Chern number $C_{12}^{\mathrm{GS}}$ vs. $\phi_3$ obtained from the scattering matrix (SM) method (solid line) and the BdG method (markers) for both the lattices in the absence and presence of gating potential $V_g$.}
        \label{fig:sm_bdg}
\end{figure}
In Fig.~\ref{fig:sm_bdg} we show the plots of the topological Chern number $C_{12}^{\mathrm{GS}}$ with respect to $\phi_3$, which are obtained using the BdG method (markers) and are compared with the SM method (solid line) when $V_g$ is absent and present at a specific value of $\eta=0.37$ (square) and $0.55$ (graphene). Clearly, in Figs.~\ref{fig:sm_bdg}(a)-(b), both methods provide remarkable agreement, except that the BdG method shows a slightly broader range for the topologically non-trivial phase with $C_{12}^{\mathrm{GS}}=\pm 1$ when gating ptential is absent ($V_g=0$), see in Fig.~\ref{fig:sm_bdg}(a). For the graphene lattice, shown in Figs.~\ref{fig:sm_bdg}(c)-(d), we observe that the BdG method shows the phase transition between the topologically trivial to non-trivial phases. However, the BdG data do not exactly fall on the SM line. Nonetheless, the values of the Chern number remain $C_{12}^{\mathrm{GS}}=\pm 1$ or $C_{12}^{\mathrm{GS}}=0$ throughout. Overall, we find that the Andreev states calculated using both methods exhibit topological phase transitions in our multiterminal Josephson junctions setup for the square and graphene lattices.   
\section{Conclusion}\label{sec:conclusion}
We have investigated the topologically non-trivial phases, arising due to the Andreev bound states, in the multiterminal Josephson junctions based on square and graphene lattices. By setting the tight-binding models system in the KWANT software~\cite{Groth2014} and employing the scattering matrix theory and the Bogoliubov-de Gennes theory in the short-junction limit, we have studied the topological nature of the Andreev bound states both in the absence and presence of an externally applied gating potential in the scattering region of MJJs. Without the gating-potential, we find that the topologically non-trivial stable phases exist with $C_{12}^{\mathrm{GS}}=\pm 1$ in a larger region of the parameter space defnined by the superconducting phase difference $\phi_3$ and the global chemical potential $\mu$. The phase transitions are marked by closing and opening of the Andreev energy bands with respect to the superconducting phase differences for suitable model parameter values. Interestingly, we also observe phase transitions between the topologically non-trivial phases upon tuning $\phi_3$ and $\mu$ for the square lattice, however, this feature is only present along $\phi_3$ in the case of graphene, as shown in Figs.~\ref{fig:chern-zerogate}(a) and~\ref{fig:chern-graphene}(a). Moreover, these non-trivial  topological phases are present even when the gating potential $V_g$ is introduced, remaining stable for moderate to large values of $V_g$ relative to $\Delta$. Additionally, for a fixed $\mu$, we observe that phase transitions also occur between the topologically non-trivial phases in response to $V_g$ for the square lattice, whereas this feature is absent along the $V_g$ for the graphene. Our results demonstrate that the topological properties can persist in experimentally relevant finite-size systems with gate potentials included.
\begin{acknowledgments}
This work was funded by the Deutsche Forschungsgemeinschaft (DFG, German Research Foundation) - 467596333. This work was partly supported by the Helmholtz Association through program NACIP.
\end{acknowledgments}
\bibliography{references_multiterminal.bib}
\end{document}